\documentclass[11pt]{article} 

\usepackage[utf8]{inputenc} 

\usepackage{geometry} 
\usepackage{graphicx} 

\usepackage[parfill]{parskip} 

\usepackage{booktabs} 
\usepackage{threeparttable}
\usepackage{color}
\usepackage{array} 
\usepackage{paralist} 
\usepackage{verbatim} 
\usepackage{subfig} 
\usepackage{url}
\usepackage{booktabs}

\usepackage[colorlinks,citecolor=blue,linkcolor=blue]{hyperref}
\usepackage{cite}
\usepackage[font=small,labelfont=bf]{caption}

\usepackage{fancyhdr} 
\usepackage{amsmath}
\usepackage{amsthm}    

	{

	\newtheorem{attack}{Attack}
	
	\newtheorem{protocol}{Protocol}
	
	}

\usepackage{amsfonts}
\usepackage{braket}

\newcommand{\basis}{b_\iterator}
\newcommand{\densebasis}{b_\denseinfocounter}
\newcommand{\basisarbitrary}{b_{\denseinfocounter+\groupcounter}}
\newcommand{\basisendgroup}{b_{\denseinfocounter+\groupsize-1}}
\newcommand{\bit}{B_\iterator}
\newcommand{\densebit}{B_\denseinfocounter}
\newcommand{\bitarbitrary}{B_{\denseinfocounter+\groupcounter}}
\newcommand{\bitendgroup}{B_{\denseinfocounter+\groupsize-1}}
\newcommand{\ciphertext}{c_\iteratortwo}
\newcommand{\nodenumber}{d}
\newcommand{\hash}{h}
\newcommand{\iterator}{i}
\newcommand{\iteratortwo}{j}
\newcommand{\cipherkey}{k_{\text{C}}}
\newcommand{\hashkey}{k_{\text{H}}}
\newcommand{\maskkey}{k_{\text{M}}}
\newcommand{\isklength}{l_k}
\newcommand{\taglength}{l_\authtag}
\newcommand{\ivlength}{l_v}
\newcommand{\authmessage}{m}
\newcommand{\authmessageiter}{\authmessage_\iterator}
\newcommand{\qubitnumber}{N}
\newcommand{\plaintext}{p_\iteratortwo}
\newcommand{\qber}{q}

\newcommand{\nonceiter}{s_\iterator}
\newcommand{\nonceitertwo}{s_\iteratortwo}
\newcommand{\numbermacvalues}{\beta}
\newcommand{\denseinfocounter}{\eta}
\newcommand{\groupcounter}{\Xi}
\newcommand{\groupsize}{\xi}
\newcommand{\authtag}{\tau}
\newcommand{\authtagiter}{\authtag_\iterator}
\newcommand{\authtagaliceiter}{\authtag_\iterator^\text{A}}
\newcommand{\authtageveiter}{\authtag_\iterator^\text{E}}
\newcommand{\authtagXiter}{\authtag_\iterator^X}
\newcommand{\authtagZiter}{\authtag_\iterator^Z}
\newcommand{\qubit}{\ket{\psi}_\iterator}
\newcommand{\densequbit}{\ket{\psi}_\denseinfocounter}
\newcommand{\qubitarbitrary}{\ket{\psi}_{\denseinfocounter+\groupcounter}}
\newcommand{\qubitendgroup}{\ket{\psi}_{\denseinfocounter+\groupsize-1}}
\newcommand{\concatenate}{||}

\title{A quantum key distribution protocol for rapid denial of service detection.}
\author{Alasdair B. Price\\ \footnotesize{Centre for Quantum Photonics and Quantum Engineering Centre for Doctoral Training,}\\ \footnotesize{H. H. Wills Physics Laboratory \& Department of Electrical and Electronic Engineering,}\\ \footnotesize{University of Bristol, Nanoscience and Quantum Information Building,}\\ \footnotesize{Tyndall Avenue, Bristol, BS8 1FD, United Kingdom}\\ \footnotesize{alasdair.price@bristol.ac.uk}\\~\\John G. Rarity and Chris Erven\\ \footnotesize{Quantum Engineering Technology Labs, H. H. Wills Physics Laboratory}\\ \footnotesize{\& Department of Electrical and Electronic Engineering, University of Bristol,}\\ \footnotesize{Nanoscience and Quantum Information Building, Tyndall Avenue,}\\ \footnotesize{Bristol, BS8 1FD, United Kingdom.}}

\begin{document}
\maketitle

\abstract{We introduce a quantum key distribution protocol designed to expose fake users that connect to Alice or Bob for the purpose of monopolising the link and denying service. It inherently resists attempts to exhaust Alice and Bob's initial shared secret, and is 100\% efficient, regardless of the number of qubits exchanged above the finite key limit. Additionally, secure key can be generated from two-photon pulses, without having to make any extra modifications. This is made possible by relaxing the security of BB84 to that of the quantum-safe block cipher used for day-to-day encryption, meaning the overall security remains unaffected for useful real-world cryptosystems such as AES-GCM being keyed with quantum devices.
}

\section{Introduction}

Quantum key distribution (QKD) enables two remote parties (Alice and Bob) to generate a shared secret, using quantum mechanics to ensure security against all possible theoretical attacks \cite{Bennett1984,Shor2000}. This means it is guaranteed quantum-safe, making BB84 (the first QKD protocol, of which there are now a number of different variants) a strong candidate for protecting future communications. Further advantage can be gained in the form of eavesdropper detection, by exploiting the disturbances introduced when an attacker measures the quantum states. Unfortunately, this opens up the potential for a denial of service (DoS) attack that can be carried out simply by increasing the error rate on the transmission line. While sometimes used as an argument against QKD \cite{CESG2016}, the risk of this happening is often overstated, as it requires an attacker to have physical access to the optical fibre, so the development of large-scale networks will mitigate any damage by enabling the quantum signal to be redirected. However, there is another way of performing DoS, that does not require an adversary to monitor all connections simultaneously, and to which all current QKD protocols are vulnerable.

\begin{figure}
	\centering
	\includegraphics[scale=0.275]{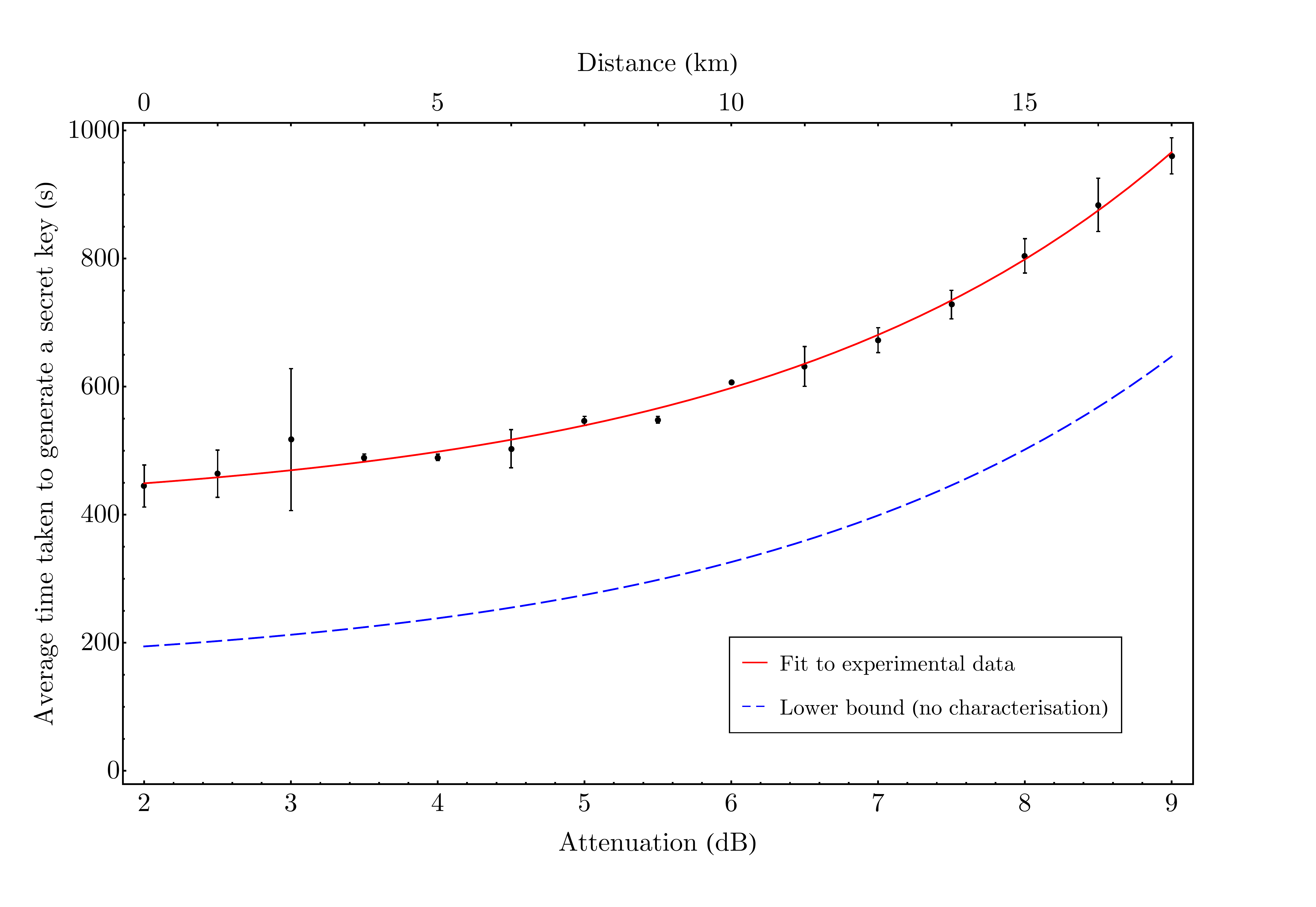}
	\caption{\label{fig:clavisplot}Time taken for a networked \emph{ID Quantique Clavis$^2$} to generate a $\sim\!10^5$-bit shared secret across a newly established connection. Each system is connected to an optical switch (introducing 1 dB of loss in each case, hence a total attenuation of 2 dB at 0 km), allowing different links to be selected. The lower bound (calculated from average secret key rates) skips device and fibre characterisation, and assumes this does not affect the performance of subsequent steps, to give the shortest possible denial of service attack duration.}
\end{figure}

To prevent man-in-the-middle attacks, it is required that the classical QKD channel be authenticated, and to retain information-theoretic security, this must be done using a Wegman-Carter message authentication code (MAC) \cite{Wegman1981} keyed with a pre-shared secret. The MAC has to be transmitted at the end of the QKD protocol, authenticating every message sent up to that point\cite{Stucki2011}, as authenticating each message individually would prohibit net positive key generation. This means neither Alice or Bob will know whether the person they are communicating with is genuine until they have a secret key, so an imposter could deny service to other users simply by opening a connection and performing QKD. Figure \ref{fig:clavisplot} shows how long this could last for, assuming only one round of key generation is carried out by the attacker. For a 10km metropolitan-area network, an \emph{ID Quantique Clavis$^2$} will communicate with an illegitimate party for roughly 10 minutes before realising! We note that the \emph{Clavis$^2$} continues to work at attenuations above 9 dB, but key generation starts to become intermittent. The average time taken for a successful round of QKD at 10 dB is close to 20 minutes, however the DoS impact could be greater if other rounds fail, which happens in over $30\%$ of cases. Ultimately, it makes sense for an attacker to maximise the attenuation on their link to keep the systems occupied for as long as possible.

In this paper, we discuss how QKD can be modified to eliminate the risk of DoS attacks that leverage provably fake users. In the real world, cryptosystems that use QKD are unlikely to employ the one-time pad in day-to-day communications. Instead, quantum-safe ciphers such as the Advanced Encryption Standard (AES) \cite{NIST2001} feature heavily\cite{Stucki2011,Sasaki2011}, as they utilise the key more efficiently. This means the overall system is not information-theoretically secure, so reducing the mathematical security of QKD in line with the encryption algorithm will not reduce the real-world security, and will actually increase it if DoS and side-channel attacks can be mitigated as a result. By making a few additional tweaks, we show that a computationally secure QKD protocol can securely generate key even from singly detected two-photon terms, and run at exactly 100\% efficiency.

\section{Preliminaries}

Authentication in QKD is traditionally performed using a Wegman-Carter MAC\cite{Wegman1981,Stucki2011,Sasaki2011}. This takes the form
\begin{equation}\label{eq:WegmanCarterMAC}
\authtag = \hash_{\hashkey}(\authmessage)\oplus \maskkey
\end{equation}
where $\hash$ is a universal hash function keyed with $\hashkey$, $\authmessage$ is the message to be authenticated (in this case, a concatenation of every transmission made over the public channel) and $\maskkey$ is the key used to mask the output of the hash. Alice calculates the tag $\authtag$ for the information she publicly announced, and sends it to Bob. He then computes the tag for the information he received, and compares it with Alice's tag. So long as the two are the same, he can be confident that the information has not come from or been modified by a third party (Eve). The same can then be done for the messages sent from Bob to Alice.

As described in the introduction, this way of handling QKD authentication creates the opportunity for an attacker to carry out a DoS attack, which we now formalise.

{\attack Eve establishes a high-loss connection with Alice and performs low bit rate QKD up to the point where she fails the authentication. During this period, Alice and Bob are unable to generate new shared keys, which may also lead to denial of service of their classical communications. The attack can be prolonged if agents of Eve are queued behind her, turning it into a distributed denial of service (DDoS) attack.\label{att:DoS}}
\\~\\
After succumbing to attack \ref{att:DoS}, Alice and Bob may find that they have exhausted their supply of pre-shared secret. This, a well-established vulnerability that also has the potential to be exploited independently (see attack \ref{att:keyexhaustion}), can be counteracted by using a post-quantum public key algorithm to authenticate the next round of QKD \cite{Roscino2016}. So long as Eve cannot break said algorithm in the short amount of time for which it is useful to her, full security is retained for all keys thereafter. However, by taking this approach, we have introduced a primitive that was not already part of the system, assuming Alice and Bob's initial secret was shared without using post-quantum cryptography. The recovery mechanism can also be triggered relatively easily, allowing attack \ref{att:keyexhaustion} to be used as a way of forcing public key algorithms to be used for every successful round of QKD. Therefore, from both simplicity and security perspectives, a reactive strategy is less than ideal.

{\attack Eve establishes a low-loss connection with Alice and performs high bit rate QKD up to the point where she fails the authentication. She or her agents repeat this until Alice no longer has enough secret key with which to construct a MAC. At this point, Alice must switch to an alternative method of key distribution to avoid indefinite denial of service.\label{att:keyexhaustion}}
\\~\\
We will later show that by careful construction of a computationally secure QKD protocol, it is possible to generate secret key from two-photon terms. This is not possible in canonical BB84 because of the following photon number splitting (PNS) attack.

{\attack Eve performs a quantum nondemolition measurement on the number of photons in each pulse. She blocks all single photon terms, and splits those containing multiple photons. She retains at least one photon in a quantum memory, and allows the remainder to carry on towards Bob. When Alice announces her preparation bases, Eve measures the stored photons, returning the same raw key as Alice (assuming zero errors). This can be sifted correctly when Bob publicly responds to Alice's original announcement.\label{att:PNS}}
\\~\\
Finally, for completeness, attack \ref{att:traditional} demonstrates what would happen if a man in the middle were able to compromise the chosen authentication scheme. As Wegman-Carter MACs are unconditionally secure, a break of this nature is not considered possible for canonical BB84, assuming Eve does not have access to the initial secret key. However, the attack will be relevant later on, when discussing the use of AES in both the QKD authentication and data encryption.

{\attack Eve intercepts the quantum bits (qubits), measures each one in a random basis and resends the results she observed in the bases she measured. She conceals this by modifying Alice's bases announcement and Bob's response, along with the authentication tags for each. Eve can now read all communications encrypted and/or authenticated using the key she shares with Alice, before forwarding them with or without modification, having re-encrypted or authenticated using the key she shares with Bob.\label{att:traditional}}

\section{The Protocol}

We begin by trying to fulfil the main objective of this paper; preventing attack \ref{att:DoS}. A trivial solution, which preserves the information-theoretic security of BB84, would be to implement some form of access control that requests Eve verify her identity before she is allowed to connect. However, if there are no further checks until the end of the protocol, this could easily be circumvented by Eve switching out Bob for herself once key generation begins. Therefore, the most sensible approach is to authenticate every message exchanged by Alice and Bob.

Ideally, this will mean modifying equation \ref{eq:WegmanCarterMAC} such that the tags can be reused without increasing the risk of an attacker being able to decrypt messages that rely on quantum keys. Brassard proposed in \cite{Brassard1983} that $\maskkey$ could be defined as the output of a random function. In practice, this can be the cipher used for the data encryption, independently keyed with $\cipherkey$, so we rewrite equation \ref{eq:WegmanCarterMAC} as

\begin{equation}\label{eq:praccompsecauth}
\authtagiter=\hash_{\hashkey}\left(\authmessageiter\right)\oplus\text{AES}_{\cipherkey}\left(\nonceiter\right)
\end{equation}

where $\nonceiter$ is a public one-time number, or ``nonce''. A number of efficient authentication schemes such as poly1305-AES \cite{Bernstein2005a}, UMAC \cite{Black2006} and VMAC \cite{Krovetz2007} take this form (though their moduli for addition vary), and their security when accompanying a known message is well established. For a 128-bit tag, all forgeries will be rejected with probability close to $1$, so long as AES cannot be distinguished from a uniform random one-to-one function, an attacker sees no more than $2^{64}$ messages and, as in conventional QKD, our hash function has small differential properties \cite{Bernstein2005b}.

As a result, just under $2^{64}$ qubits can be individually accompanied by a MAC, assuming Bob uses a separate initial secret key with an independent nonce for sending authenticated replies to Alice. A small number of tags must also be retained for messages relating to other parts of the protocol, such as error correction.

For finite key security, $\Omega\left(10^5\right)$ raw bits must be exchanged and processed \cite{Scarani2008}, meaning we can complete $\mathcal{O}\left(10^{14}\right)$ rounds of QKD before the scheme \emph{needs} to be rekeyed. The impact of this is two-fold. First, attack \ref{att:keyexhaustion} is no longer viable, as an eavesdropper needs to establish more than eighteen billion billion connections before Alice and Bob will be prevented from constructing any more MACs of the form given by equation \ref{eq:praccompsecauth}. Second, even if Eve were able to ensure key generation only failed at the very last moment, the number of times she would have to repeat her attack in order to exhaust Alice and Bob's shared secret is still on the order of a hundred trillion, given the rekeying limit specified above, and assuming they only began with the minimum number of bits required to construct a secure MAC. For networks of sufficient size, we would expect them to find a link that she cannot influence long before reaching that limit.

We note that the choice to use AES-256 for both data encryption and QKD authentication is not just for the sake of simplicity, or so we can be confident our cryptosystem remains quantum-safe (although as this is our reason for using QKD in the first place, it is obviously important). Suppose that, despite all the analysis that has taken place up to this point, AES has an undisclosed flaw that allows attack \ref{att:traditional} to be carried out by a select few. The result would be catastrophic. However, it would be no different compared to if the AES-based data encrypter had been paired with canonical BB84 instead, because the encryption can be broken directly in either case, meaning attack \ref{att:traditional} offers no advantage. Of course, the chances of this happening are thought to be very low, and so even if the one-time pad were used for data encryption, the comparative reduction in mathematical security is outweighed by increased resilience against DoS attacks.

In a world where Eve cannot compromise AES, she may carry out an unsuccessful version of attack \ref{att:traditional} on only some of the qubits. Although Alice and Bob will be aware of her presence, there would be no way of knowing which qubits had been targeted in standard BB84, so the entire protocol would have to be aborted. In our case, the individual authentication of every basis would allow Alice and Bob to identify which qubits had been attacked in this way, giving them the option to keep those that were unaffected.

The above changes ensure that, if Eve tries to carry out attack \ref{att:DoS}, she will deny service for fractions of seconds rather than tens of minutes before her presence becomes obvious. This is achieved without a reduction in the mathematical security of real-world QKD-based cryptosystems. The next step is to look at whether we can gain any further benefits by capitalising on our use of a computationally secure MAC.

Now that every basis announcement is accompanied by an authentication tag, an interesting property emerges. There are only two possible tags for any given key/nonce pair, depending on whether the qubit was prepared in the $X$ basis or the $Z$ basis (though the exact values are unpredictable for anyone not in possession of the key). This means that if Alice decides to send the tags on their own, without the plaintext basis announcement that they authenticate, Bob can work out how he should have measured the qubit, by comparing the tags he would expect for each option. The bound for rejecting forgeries will remain the same, as the plaintext can always be ignored in the case where the bases are publicly announced.

Ideally, lack of knowledge about Alice and Bob's shared secret will prevent Eve from also identifying the correct bases using the authentication tags. That is, if they provide confidentiality, which is not a traditional requirement of a MAC, then she will no longer be able to carry out attack \ref{att:PNS}. This can easily be shown to be true for tags of the form given in equation \ref{eq:praccompsecauth}.

AES-CTR (AES running in counter mode \cite{Dworkin2001}) encrypts a plaintext message, $\plaintext$, as follows

\begin{equation}\label{eq:AESCTR}
\ciphertext= \plaintext \oplus \text{AES}_{\cipherkey}(\nonceitertwo)
\end{equation}

where $\ciphertext$ is the ciphertext, and $\nonceitertwo$ contains a counter that increments with each value of $\iteratortwo$, never repeating for any given $\cipherkey$. So long as the counter is of length 64 bits or more (with the remainder of $\nonceitertwo$ comprising of random bits), $2^{64}$ messages can be sent with only a minimal chance of Eve being able to recover the plaintext \cite{Bellare1997}.

We observe that if one were to set $\plaintext=\hash_{\hashkey}\left(\authmessageiter\right)$, then equations \ref{eq:praccompsecauth} and \ref{eq:AESCTR} become the same. This means an AES-based MAC of Wegman-Carter form can be viewed as an implementation of AES-CTR, and so Eve will be unable to work out which basis to measure in, given only a properly implemented 128-bit tag.

From the above, we have established that transmitting the basis information as proposed means two-photon pulses can contribute to the secure key rate. However, it is still possible to implement an alternative method for PNS, on higher-order multiphoton terms (see attack \ref{att:PNSnew}). All protocols are vulnerable to this unless, as in \cite{Lo2005a} and \cite{Stucki2005}, additional eavesdropper detection mechanisms are in place.

{\attack Eve performs a quantum nondemolition measurement on the number of photons in each pulse. She blocks all single and two-photon terms, but splits those containing three or more photons. She retains at least two photons in a quantum memory, and allows the remainder to carry on towards Bob. Eve then performs unambiguous state discrimination \emph{\cite{Enk2002}} on the qubits in her possession and returns a proportion of Alice's raw key dependent on the number of photons she split out of each pulse.\label{att:PNSnew}}
\\~\\
Of course, if the tags provide a level of confidentiality sufficient to prevent attack \ref{att:PNS}, there is no longer any reason for them to be transmitted after Bob has measured the qubits, as Eve is unable to obtain the information required to perform a man-in-the-middle attack. If the tags are transmitted in advance, Bob can work out how he needs to measure before each qubit arrives, increasing the efficiency of the protocol from 50\% to 100\%.

As a result, our protocol does not need to be implemented using biased bases which, conditional on the number of photons transmitted, are used to asymptotically double the efficiency of BB84 \cite{Lo2005b}. In fact, given we have already waived our interest in information-theoretic security, transmitting the tags in advance of the qubits is a slightly preferable solution. This is partly because the efficiencies of real and simulated biased basis experiments are still noticeably lower than 100\% \cite{Erven2009,Wei2013}, however assuming no additional countermeasures are employed, the protocol described in \cite{Lo2005b} is also vulnerable to a more simplistic PNS attack than that which is applicable to vanilla BB84. This, attack \ref{att:PNSnomemory}, is possible due to the recommendation that key be generated from a single basis, with the other used only for eavesdropper detection. The fact a quantum memory is no longer required makes it a much more realistic exploit for modern-day implementations than attack \ref{att:PNS}, emphasising why it is imperative to use decoy states in any current system relying on biased bases. In contrast, the aforementioned \emph{Clavis$^2$} predominantly uses unbiased SARG04 \cite{Scarani2004}, which has the same level of PNS-resistance as the protocol described herein, and falls back on unbiased BB84 for short distances, where SARG04 is not proven secure \cite{Branciard2005}. This may be considered acceptable so long as quantum memories remain in the early stages of development.

\begin{figure}
	\includegraphics[scale=0.23]{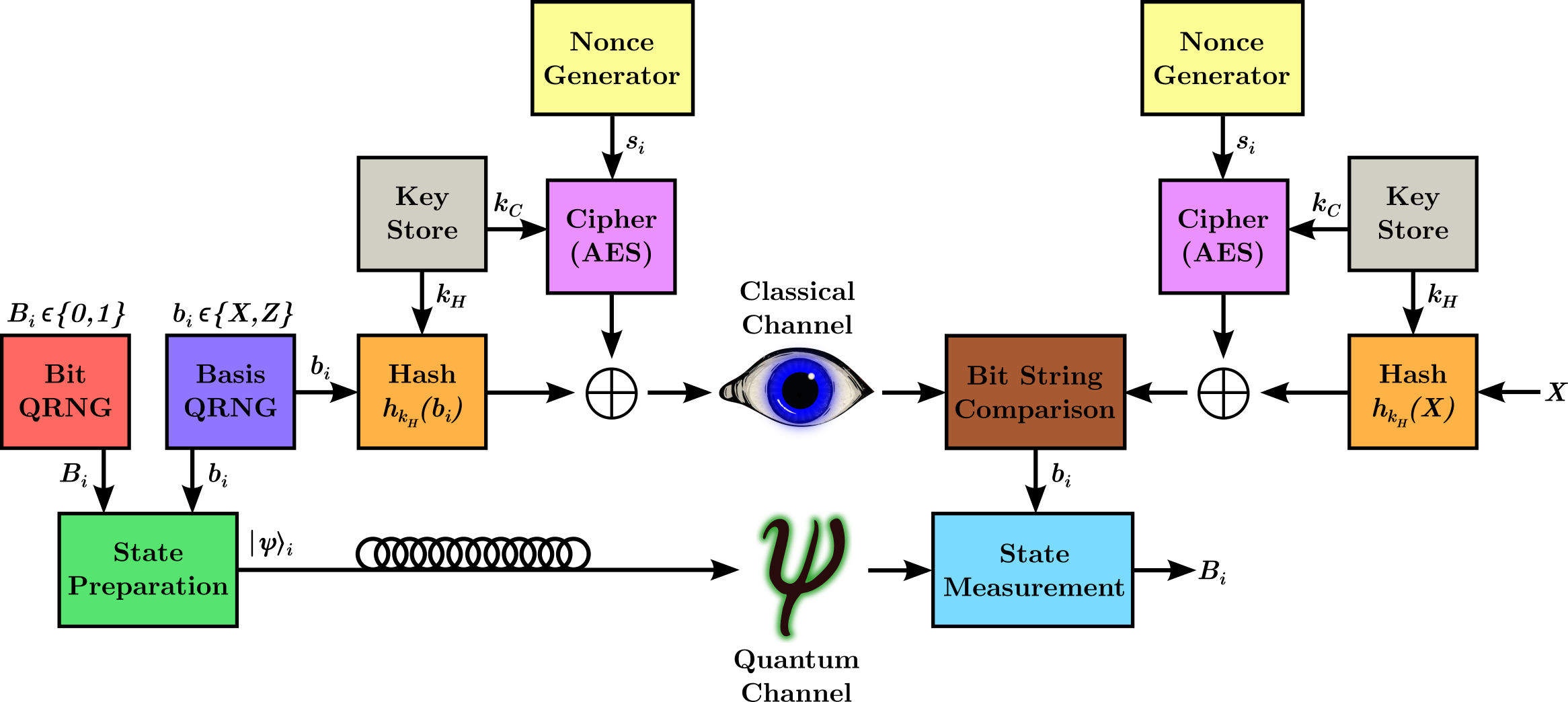}
	\caption{\label{fig:protocol} Block diagram showing the transmission of a single bit of key from Alice to Bob, as part of BB84-AES in its reduced processing form.}
\end{figure}

{\attack Assume Eve does not possess a quantum memory, but is otherwise unchanged. She performs a quantum nondemolition measurement on the number of photons in each pulse and blocks all single photon terms. For the remainder, she splits off at least one photon from every pulse, and allows at least one photon to carry on towards Bob. Eve immediately measures her copy in the key generation basis. When Alice and Bob publicly sift their qubits, she can identify those used for eavesdropper detection, and discard any information she has on them. Every bit of her final key has now been correctly measured, without revealing her presence.\label{att:PNSnomemory}}
\\~\\
Protocol \ref{prot:basic} pulls together the methods we have developed for performing computationally secure, but still quantum-safe, QKD. A streamlined version is presented in figure \ref{fig:protocol}, the details of which can be found in the next section. Up until now, we have focused solely on utilising AES, because of its ubiquity in modern communications, and position as the de facto quantum-safe alternative to the one-time pad. However, should AES ever become compromised in some way, it would be trivial to substitute in an alternative cipher (the post-quantum security of Serpent-256 is currently under evaluation \cite{Augot2015}, for example).

While we have assumed the quantum key will be used in computationally secure cryptosystems, it is still sensible to investigate the impact of a user who insists on encrypting their data with the one-time pad in a bespoke setting, despite its low efficiency and lack of authenticated encryption modes. In this scenario, we retain the advantages of our protocol, but also acquire everlasting security \cite{Mosca2013} (the plaintext cannot be recovered from the information available to Eve if she develops unlimited computational power after key exchange is complete). This, along with perfect forward secrecy (previously generated keys will be unaffected if the initial shared secret has not been refreshed and the current round of the protocol becomes compromised), cannot be achieved if the key is encrypted directly with AES. For such a scheme, perfect forward secrecy is unattainable because anyone in possession of the long-term secret can use it to extract past session keys from the ciphertexts, rather than returning a set of bases that are no longer of any use. Similarly, compromising a previous shared secret at a later date will expose all keys distributed thereafter, even if the secret is updated after every key exchange with material from that session. Therefore, one should take care not to be fooled into thinking direct encryption of the key is a valid simplification of our protocol. Of course, a system based on this would not provide eavesdropper detection either.

\begin{table*}[h]
	\centering
	\small
		\begin{tabular}{p{15cm}}
			\toprule[2\heavyrulewidth]
			\vspace{-12pt}
			{\protocol \emph{BB84-AES (basic version)}\label{prot:basic}} \\
			\midrule
			SUMMARY: Alice expands a shared secret with Bob, using computationally secure QKD and quantum-safe primitives.       
			\begin{enumerate}
				\item \emph{One-Time Setup.}
				\begin{enumerate}
					\item An $\isklength$-bit secret is shared between Alice and Bob using out-of-band communications, a trusted third party or a post-quantum public key algorithm.
					\item An $\ivlength$-bit initialisation vector is transmitted from Alice to Bob in the clear, where $\ivlength\leq 64$.
				\end{enumerate}
				\item \emph{Nonce Generation.} A single-use number $\nonceiter$ is constructed by appending a $(128-\ivlength)$-bit counter to the initialisation vector. The counter starts at $0$ and increments after each call made to the generator. It must be maintained across all rounds of QKD that use the same initial shared secret, and is not to be confused with the index $\iterator$ used in the mathematics of this paper, where $1\leq\iterator\leq\qubitnumber$.
				\item \emph{Authentication Tags.}
				\begin{enumerate}
					\item The shared secret is split into a $256$-bit cipher key, $\cipherkey$, and an $(\isklength-256)$-bit hash key, $\hashkey$.
					\item Alice generates a cryptographically secure random number, which is used to select a basis $\basis\in\{X,Z\}$, and computes the tag $\authtagaliceiter=\hash_{\hashkey}\left(\basis\right)\oplus\text{AES}_{\cipherkey}\left(\nonceiter\right)$. $\hash$ is a universal hash function, the output of which can be called from memory after it has been evaluated once for each basis, and AES is the Advanced Encryption Standard block cipher.\label{step:taggen}
					\item Bob calculates $\authtagXiter=\hash_{\hashkey}\left(X\right)\oplus\text{AES}_{\cipherkey}\left(\nonceiter\right)$ and $\authtagZiter=\hash_{\hashkey}\left(Z\right)\oplus\text{AES}_{\cipherkey}\left(\nonceiter\right)$.\label{step:chktaggen}
				\end{enumerate}
				\item \emph{Key Exchange.}\label{step:keyexchange}
				\begin{enumerate}
					\item Alice prepares a qubit $\qubit$ by generating a cryptographically secure random number, $\bit\in\{0,1\}$, and encoding it in the basis $\basis$.\label{step:prepqubit}
					\item Alice sends $\authtagaliceiter$ to Bob, closely followed by $\qubit$.\label{step:sendqubit}
					\item Bob compares $\authtagaliceiter$ with $\authtagXiter$ and $\authtagZiter$, to identify the basis in which he should measure. Upon receipt of $\qubit$, he will return $\bit$ with probability $100\%-\qber$, where $\qber$ is the quantum bit error rate.\label{step:findbasis}
					\item Bob announces whether or not the qubit arrived, by means of an authenticated response. He should maintain a separate nonce generator to Alice, paired with a different shared secret. As Bob's response need only be ``Yes'' or ``No'', he may choose to transmit it in the same way as Alice sends her bases.\label{step:bobresponse}
				\end{enumerate}
				\item \emph{Loop.} Steps \ref{step:taggen}, \ref{step:chktaggen} and \ref{step:keyexchange} are repeated for the remaining $\qubitnumber-\iterator$ qubits sent from Alice to Bob. As multiple tags can be constructed in parallel, this may begin prior to completion of the previous iteration. \label{step:rinserepeat}
				\item \emph{Post Processing.}
				\begin{enumerate}
					\item Error correction and privacy amplification are carried out as in BB84. The messages sent during this step can be authenticated in the same way as above.
				 	\item $\isklength$ bits are taken from the final key and stored for use as the initial secret in the next round of QKD, and a new initialisation vector is publicly agreed upon.
				\end{enumerate}
			\end{enumerate}\\
			\bottomrule[2\heavyrulewidth]
		\end{tabular}
\end{table*}

\section{Optimisations}

While it is perfectly feasible to implement protocol \ref{prot:basic} as presented herein, there are a number of changes that can be made to reduce demand on the computational and/or communications resources. The first of these is summarised in protocol \ref{prot:redproc}, where we allow Bob to check only whether the tag he receives is a match for that corresponding to a measurement in the $X$ basis. This requires marginally less memory and processing time than individual basis authentication in otherwise-standard BB84. The trade-off is that if Eve measures in the $Z$ basis, she no longer needs to be able to forge the corresponding authentication tag, ensuring only that the one she forwards, $\authtageveiter$, is different to that sent by Alice. However, Eve still has not broken the authentication scheme (she cannot obtain any basis information or force Bob to measure in the $X$ basis), and so this kind of interference will be exposed by the quantum bit error rate (QBER). Table \ref{tab:authmodify} gives the outcomes for all of Eve's possible strategies. It is clear that $\authtageveiter\equiv\authtagaliceiter$ remains optimal.

\begin{table*}[h]
	\centering
	\small
	\begin{threeparttable}
		\begin{tabular}{p{15cm}}
			\toprule[2\heavyrulewidth]
			\vspace{-6pt}
			{\protocol \emph{BB84-AES (reduced processing)}\label{prot:redproc}} \\
			\midrule
			SUMMARY: Replaces steps \ref{step:chktaggen} and \ref{step:findbasis} in protocol \ref{prot:basic}, halving the number of XOR operations and tag comparisons that Bob has to carry out.     
			\begin{enumerate}
				\setcounter{enumi}{2}
				\item \emph{Authentication Tags.}
				\begin{enumerate}
					\setcounter{enumii}{2}
					\item Bob calculates $\authtagXiter=\hash_{\hashkey}\left(X\right)\oplus\text{AES}_{\cipherkey}\left(\nonceiter\right)$.
				\end{enumerate}
			\item \emph{Key Exchange.}
				\begin{enumerate}
				\setcounter{enumii}{2}
				\item Bob compares $\authtagaliceiter$ with $\authtagXiter$. If it matches, he will choose to measure in the $X$ basis. Otherwise, he will choose to measure in the $Z$ basis. Upon receipt of $\qubit$, he will return $\bit$ with probability $100\%-\qber$, where $\qber$ is the quantum bit error rate.
				\end{enumerate}
			\end{enumerate}\\
			\bottomrule[2\heavyrulewidth]
		\end{tabular}
		\begin{tablenotes}[flushleft]
			\item {\bf Can be combined with: }BB84-AES (reduced bandwidth)
		\end{tablenotes}
	\end{threeparttable}
\end{table*}

\begin{table*}
	\centering
	\small
	\caption{\label{tab:authmodify}Showing the probability of a bit-flip error occurring between Alice and Bob depending both on the bases chosen by each of the three parties, and whether or not Eve blindly modifies the authentication tag.}
		\begin{tabular}{ccccc}
			\toprule[2\heavyrulewidth]
			Alice's Basis & Eve's Basis & Forwarding Choice & Bob's Basis & Prob(error) \\
			\midrule
			X             & X                  &   $\authtageveiter=\authtagaliceiter$   & X           & 0           \\
			X             & X                  &   $\authtageveiter\neq \authtagaliceiter$  & Z           & 0.5         \\
			X             & Z                  &  $\authtageveiter=\authtagaliceiter$   & X           & 0.5         \\
			X             & Z                  &   $\authtageveiter\neq \authtagaliceiter$   & Z           & 0.5         \\
			Z             & X                  &   $\authtageveiter=\authtagaliceiter$   & Z           & 0.5         \\
			Z             & X                  &   $\authtageveiter\neq \authtagaliceiter$   & Z           & 0.5         \\
			Z             & Z                  &  $\authtageveiter=\authtagaliceiter$    & Z           & 0           \\
			Z             & Z                  &   $\authtageveiter\neq \authtagaliceiter$   & Z           & 0     \\     
			\bottomrule[2\heavyrulewidth]
		\end{tabular}
\end{table*}

Next, we look at the effect of requiring the classical channel to transmit $128\times$ the number of bits transferred over the quantum channel. Given the \emph{Clavis}$^2$ emits laser pulses clocked at 5 MHz \cite{IDQ2013}, the classical data rate needs to be 640 Mbit/s. For comparison, the Bristol and UK quantum networks on which the \emph{Clavis}$^2$ systems are being deployed, both have SFP+ and QSFP+ channels with capacities of 10 Gbit/s and 40 Gbit/s respectively. While the gap appears large between what we need and what we can provide, pre-commercial quantum hardware has been shown to be capable of reaching super-GHz clock speeds \cite{Sibson2017}. Due to the way in which the states were encoded in this example, the actual clock rate of BB84 was only 560 MHz, however to avoid a potential future where our protocol necessitates two transceivers be multiplexed together, we can reduce our tag lengths as described in protocol \ref{prot:redband}. This remains secure for up to $2^{32}$ messages \cite{Bernstein2005b}, allowing $\mathcal{O}\left(10^{4}\right)$ full rounds of QKD per initial key, and brings the classical communications requirements to within the capabilities of QSFP28 or CFP4 transceivers.

\begin{table*}[h]
	\centering
	\small
	\begin{threeparttable}
		\begin{tabular}{p{15cm}}
			\toprule[2\heavyrulewidth]
			\vspace{-6pt}
			{\protocol \emph{BB84-AES (reduced bandwidth)}\label{prot:redband}} \\
			\midrule
			SUMMARY: Replaces the 128-bit tags in protocol \ref{prot:basic} with 64-bit tags of the same form. UMAC \cite{Black2006} and VMAC \cite{Krovetz2007} both provide such functionality, without dropping below the required security level.
			\\
			\\
			\bottomrule[2\heavyrulewidth]
		\end{tabular}
		\begin{tablenotes}[flushleft]
			\item {\bf Can be combined with: }BB84-AES (reduced processing), BB84-AES (dense information transfer)
		\end{tablenotes}
	\end{threeparttable}
\end{table*}

The final optimisation reduces demand on the classical channel by grouping multiple bases into a single authentication tag (protocol \ref{prot:denintrans}). The time taken to establish the presence of a fake user should not change significantly, because the tags are still transmitted ahead of the first qubit in every group. Of course, the processing at Bob's end will be expected to take slightly longer than before, as a MAC that represents $\groupsize$ bases will have $\numbermacvalues=2^\groupsize$ possible values for each key/nonce pair. His method for identifying the correct set of measurements differs from protocol \ref{prot:basic} in that he must compute all possible hashes and store them in a lookup table. He can then XOR the incoming tag with the AES-generated key, and compare. Combining protocol \ref{prot:redband} with protocol \ref{prot:denintrans} will speed up the hash function \cite{Krovetz2007}, thereby reducing the time taken to construct the table. The necessary calculations can be performed during downtime, or in parallel with device and fibre characterisation, or in parallel with a previous round of QKD provided each initial shared secret is used across multiple rounds. An important subtlety, that is also true for protocols \ref{prot:basic}, \ref{prot:redproc} and \ref{prot:redband}, is the hashes only need to be computed once so long as the initial secret remains unchanged, meaning that until this is refreshed, the lookup table does not need to be reconstructed.

To prevent a simple timing attack, Alice can never send the qubits until the worst-case lookup time has elapsed, so Bob must take care to select a search algorithm that is optimal in this regard, such as binary search \cite{Knuth1998} which makes no more than $\lfloor\log_2 \numbermacvalues\rfloor +1 = \groupsize +1$ comparisons.

\begin{figure}
	\centering
	\includegraphics[scale=0.25]{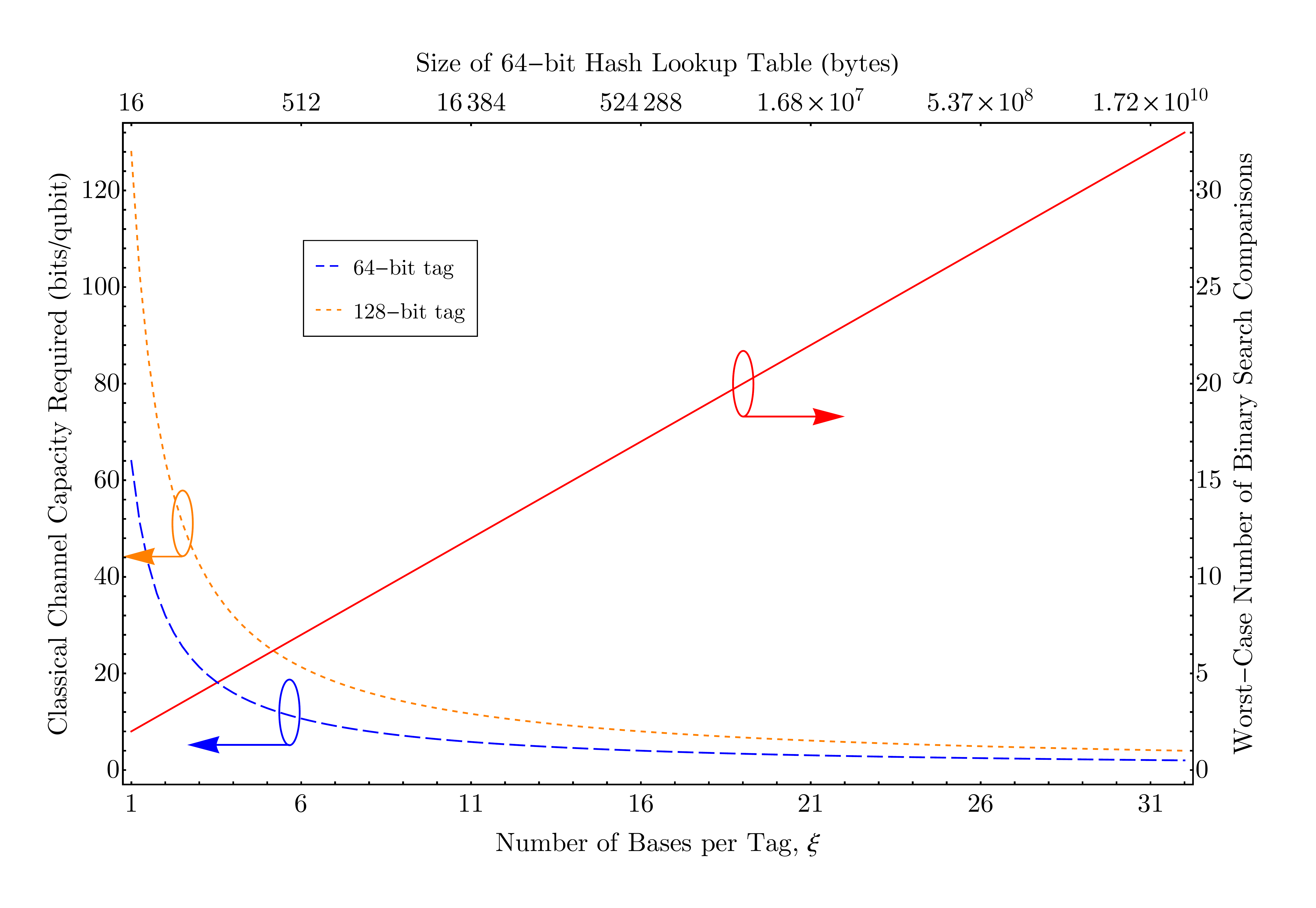}
	\caption{\label{fig:commsvscomputationplot}Illustrating how changing the number of bases represented by a single authentication tag affects both classical communication and computational resource requirements. To get a rough estimate for how our protocol will perform on a particular physical system, one can multiply the classical channel capacity by the QKD clock rate, and worst-case number of comparisons by the time taken to perform a single binary search comparison. The size of a 128-bit hash lookup table will always be double that of its 64-bit counterpart.}
\end{figure}

The exact value of $\groupsize$ reflects a trade-off between computational and communications resources, and it is clear from figure \ref{fig:commsvscomputationplot} that the greatest benefits can be achieved when $1<\groupsize\ll32$, because of the exponential behaviour of both classical channel capacity and memory requirements. As a concrete example, we will consider the Bristol quantum network, which is hosted on pre-existing infrastructure, with each node's server containing 64 Intel Xeon E5-2697A v4 processors. By implementing a binary search on a single CPU, without hardware-specific optimisation, we can estimate the performance of our protocol on a real system. If we assume a 64-bit tag and want to employ only a single SFP+ (QSFP+) channel, then $\groupsize=8$ ($\groupsize=2$) maximises the QKD clock rate while trying to use the least possible memory. In this case, it takes $6.940 \pm 0.085$ ns ($2.085 \pm 0.017$ ns) to run the search, allowing for a $1.153 \pm 0.014$ GHz ($0.959 \pm 0.008$ GHz) clock and consuming 2048 bytes (32 bytes) of memory, out of 87.7 GiB available and 131.7 GiB total RAM. To run a hypothetical 1.72 GHz-clock BB84 device based on the technology in \cite{Sibson2017} would require $\groupsize=12$ ($\groupsize=3$). In this instance, the search takes $9.692 \pm 0.039$ ns ($2.881 \pm 0.036$ ns), and 32,768 bytes (64 bytes) of memory is required. However, it is important to note that while these parameters are sufficient to enable the use of presently-installed transceivers, the quantum clock is still capped at $1.238 \pm 0.005$ GHz ($1.041 \pm 0.013$ GHz) because of the maximum search time. Hence, some parallelisation will also be required, in that each search must begin before the previous one is guaranteed to have finished.

Technically, the higher the value of $\groupsize$, the easier it is for Eve to guess one of the $2^\groupsize-1$ other authentication tags that Bob will accept. A correct guess is still highly improbable, and so she will almost certainly be detected, however even if successful, Eve controls only whether or not Bob measures with the same bases as Alice. Hence, this is nothing more than a restricted version of the strategy she can employ in protocol \ref{prot:redproc} and, in the unlikely case of an odds-defying set of forgeries, Alice and Bob will be made aware of Eve's presence by the QBER.

\begin{table*}[h]
	\centering
	\small
	\begin{threeparttable}
		\begin{tabular}{p{15cm}}
			\toprule[2\heavyrulewidth]
			\vspace{-6pt}
			{\protocol \emph{BB84-AES (dense information transfer)}\label{prot:denintrans}} \\
			\midrule
			SUMMARY: Replaces steps \ref{step:taggen}, \ref{step:chktaggen}, \ref{step:prepqubit}, \ref{step:sendqubit}, \ref{step:findbasis} and \ref{step:rinserepeat} in protocol \ref{prot:basic}, grouping multiple bases into a single tag to reduce the necessary channel capacity by a factor of $\taglength(\groupsize-1)$. $\taglength$ is the tag length in bits, and $\groupsize$ is the number of bases per tag. We redefine the range of $\iterator$ such that $1\leq\iterator\leq\frac{\qubitnumber}{\groupsize}$.
			\begin{enumerate}
				\setcounter{enumi}{2}
				\item \emph{Authentication Tags.}
				\begin{enumerate}
					\setcounter{enumii}{1}
					\item Alice generates $\groupsize$ cryptographically secure random numbers, which are used to select bases $\densebasis$ through $\basisendgroup$, where $\basisarbitrary\in\{X,Z\}$, $\denseinfocounter=1+(\iterator-1)\groupsize$ and $\groupcounter\in\{0,\ldots,\groupsize-1\}$. It is required that $1<\groupsize\ll\qubitnumber$. She computes the tag $\authtagaliceiter=\hash_{\hashkey}\left(\densebasis\concatenate\ldots\concatenate\basisendgroup\right)\oplus\text{AES}_{\cipherkey}\left(\nonceiter\right)$. $\hash$ is a universal hash function, AES is the Advanced Encryption Standard block cipher, and $\concatenate$ is used to indicate a concatenation.\label{step:densetaggen}
					\item Bob calculates $\hash_{\hashkey}\left(\densebasis\concatenate\ldots\concatenate\basisendgroup\right)$ for all $2^\groupsize$ possible values of $\densebasis\concatenate\ldots\concatenate\basisendgroup$, storing the results in ascending order. He also evaluates $\text{AES}_{\cipherkey}\left(\nonceiter\right)$ separately. \label{step:lookupconstruct}
				\end{enumerate}
			\item \emph{Key Exchange.}\label{step:densekeyexchange}
				\begin{enumerate}
				\item Alice prepares the qubits $\densequbit$ to $\qubitendgroup$. This is done by generating $\groupsize$ cryptographically secure random numbers $\densebit$ through $\bitendgroup$, where $\bitarbitrary\in\{0,1\}$, and encoding them in the bases $\densebasis$ through $\basisendgroup$ respectively.
				\item Alice sends $\authtagaliceiter$ to Bob, closely followed by all $\qubitarbitrary$ for the corresponding value of $\iterator$.
				\item Bob computes $\authtagaliceiter\oplus\text{AES}_{\cipherkey}\left(\nonceiter\right)$ and checks it against the lookup table he constructed in step \ref{step:lookupconstruct}, to identify the bases in which he should measure. Upon receipt of $\qubitarbitrary$, he will return $\bitarbitrary$ with probability $100\%-\qber$, where $\qber$ is the quantum bit error rate.
				\end{enumerate}
			\item \emph{Loop.} Steps \ref{step:densetaggen}, \ref{step:lookupconstruct} and \ref{step:densekeyexchange} are repeated for the remaining $\qubitnumber-\iterator\groupsize$ qubits sent from Alice to Bob. As multiple tags can be constructed in parallel, this may begin prior to completion of the previous iteration.
			\end{enumerate}\\
			\bottomrule[2\heavyrulewidth]
		\end{tabular}
		\begin{tablenotes}[flushleft]
			\item {\bf Can be combined with: }BB84-AES (reduced bandwidth)
		\end{tablenotes}
	\end{threeparttable}
\end{table*}

\section{Conclusion}

We have shown that, by reducing the mathematical security of BB84, it is possible to almost instantly detect denial of service that leverages fake users, something which no other quantum key distribution protocol has been shown to be capable of. Our design is inherently resilient against attacks that aim to exhaust Alice and Bob's supply of initial secret key, but does not lead to large memory overheads because of this, nor does it operate reactively by falling back on public key cryptography. In changing how and when the bases are announced, we are able to achieve exactly 100\% efficiency and, instead of posing a risk to security, two-photon terms now contribute positively to the final key rate, independently of the distance or number of bits exchanged, and without any further cost.

Such advantages are possible only so long as the output of the cipher used to construct our authenticators is indistinguishable from the output of a random permutation.  This criterion is the same as that for ensuring the security of quantum-safe encryption schemes used in day-to-day communications, so having to sacrifice information-theoretic security is not overly concerning. At any rate, the chance that the above assumption will be violated is far lower than the likelihood of an attacker exploiting one of the weaknesses that our protocol defends against. If one were to insist on unconditional security, individual basis authentication could be performed using AES tags in standard BB84, reauthenticating everything at the end with a traditional Wegman-Carter MAC. However, attack vectors may still exist for exhausting the initial shared secret and, given the issues we have raised over implementing biased bases without the necessary hardware for decoy states, BB84-AES remains preferable, particularly for minimilistic implementations and retrofitting systems already in the field.

A final novelty of our protocol is that, by daisy-chaining multiple Alice/Bob pairs, it is possible to supply an arbitrary amount of quantum-safe quantum randomness with everlasting security to someone who cannot directly access a node containing a quantum random number generator (QRNG). Of course, the resource requirements scale badly (for a chain of $\nodenumber$ nodes, the QRNG would need to generate $2^{\nodenumber-1}$ bit strings) and while the idea may be academically interesting, it is unclear whether such functionality is of any real-world use.

Adapting our work for the Six State Protocol \cite{BechmannPasquinucci1999} (which we call SSP-AES) and BBM92 \cite{Bennett1992a} (likewise, BBM92-AES) is trivial, however the ease with which it can be applied to other forms of quantum key distribution is less well-defined. An advantage can certainly be gained by incorporating decoy states, although this is yet to be quantified. We have shown that the intersection between modern and quantum cryptography should be explored in more detail, with greater collaboration between researchers on both sides, as this area still seems largely untapped and ripe for real-world improvements in algorithms and implementations.

\section*{Acknowledgements}

A. B. Price. was supported by the Bristol Quantum Engineering Centre for Doctoral Training, EPSRC grant EP/L015730/1. The authors acknowledge the UK Quantum Technology Hub for Quantum Communications Technologies, EPSRC grant EP/M013472/1. Thanks also go to K. G. Paterson and D. L. D. Lowndes for useful conversations.


\bibliographystyle{abbrv}

\bibliography{QKD_Without_Sifting} 



\end{document}